\documentclass{article}
\usepackage[dvipdfmx]{graphicx}
\usepackage{authblk}
\usepackage[version=3]{mhchem}
\usepackage{wrapfig}
\usepackage{cite}
\begin{document}

\title{
  \LARGE{Recent status of FPCCD vertex detector R\&D} \vspace{0.3cm}
  \\  \small{"Talk presented at the International Workshop on Future Linear
  Colliders (LCWS15), 2-6 November 2015, Whistler, Canada.''}
}
\author[1]{Shunsuke Murai}
\author[1]{Akimasa Ishikawa}
\author[1]{Tomoyuki Sanuki}
\author[2]{Akiya Miyamoto}
\author[2]{Yasuhiro Sugimoto}
\author[2]{Clancha Constantino}
\author[3]{Hisao Sato}
\author[4]{Hirokazu Ikeda}
\author[1]{Hitoshi Yamamoto}

\affil[1]{Tohoku University}
\affil[2]{KEK}
\affil[3]{Shinshu University}
\affil[4]{JAXA}

\date{}
\maketitle

\begin{abstract}
The Fine Pixel CCD (FPCCD) is one of the candidate sensor technologies for the ILC vertex detector. It will be located near interaction point and require high radiation tolerance. It will thus be operated at -40$^\circ$C to improve radiation tolerance. In this paper, we report on the status of neutron radiation tests, on a cooling system using two-phase \ce{CO2} with a gas compressor for circulation, and on the mechanical structure of the FPCCD ladders.
\end{abstract}

\section{Introduction}
The main role of a vertex detector in ILC is to identify b-quark and c-quark from light quarks and gluones. In general, a b-jet has 2 vertices and c-jet has 3 vertices, while light quarks and gluones have 1 vertex. A vertex detector uses that information to identify quarks. Since lifetime of b-quark and c-quark is very short with about 1 pico second, the requirement for the impact parameter resolution is $5\oplus10/(p\beta\sin^{3/2}{\theta})\, \mathrm{({\mu}m)}$\cite{ref1}. The innermost layer is located at radius of 1.6cm from the beamline for good impact parameter resolution, thus it is exposed to many $e^+e^-$ backgrounds from beam-beam interaction. Hit occupancy less than a few \% is necessary for track reconstruction but it would be  about 10 \% using a vertex detector which has normal pixel size $25\times25 \mathrm{({\mu}m)}$ when it accumulates all the hits from one beam train. \par
There are two solutions to get a low pixel occupancy. One is to read out many times in one beam train. Its problem is a EMI noise from beam. Another is to use a small pixel size as $5\times5 \mathrm{({\mu}m)}$ and to read out during the train gap. In this way, there are no EMI noise. The Fine Pixel CCD (FPCCD) vertex detector adopts the second way\cite{ref2}.

\section{Neutron irradiation test}
CCD is generally sensitive to radiation damage especially from heavy particles. Since neutrons would come from beam dump in ILC, neutron tolerance should be checked.  A neutron irradiation test was performed at CYRIC of Tohoku University from 15th to 17th Oct. 2014. The energy of neutron beam produced by 70 MeV proton beam through a reaction of $Li+p{\rightarrow}Be+n$ is about 65 MeV. An FPCCD prototype whose pixel size is $6\times6\mathrm{({\mu}m)}$ was irradiated 2 hours and its fluence was $1.78\times10^{10}\mathrm{(n_{eq}/cm^{2})}$. In ILC experiment neutron fluence is estimated at $1.85\times10^{9}\mathrm{(n_{eq}/cm^{2}/year)}$\cite{neutron} thus the neutron fluence coresponds to 19 years of $\sqrt{s}=500 \mathrm{GeV}$ ILC beam time shared by two detectors. After the irradiation, the FPCCD prototype was kept at room temperature to see its anealing effect. 

\section{Measurement of performance}
We have studied three parameters to measure the radiation tolerance of the FPCCD. They are average dark charge of all pixels, hot pixel fraction and charge transfer inefficiency. These parameters are measured 3, 9, 23 and 199 days after irradiation to see annealing effect. Status of these studies are reported in the following\cite{ref3}.

\subsection{Dark charge}
\begin{figure}
	\begin{tabular}{cc}
	\begin{minipage}[t]{0.45\hsize}
   \centering
   \includegraphics[width=5.5cm]{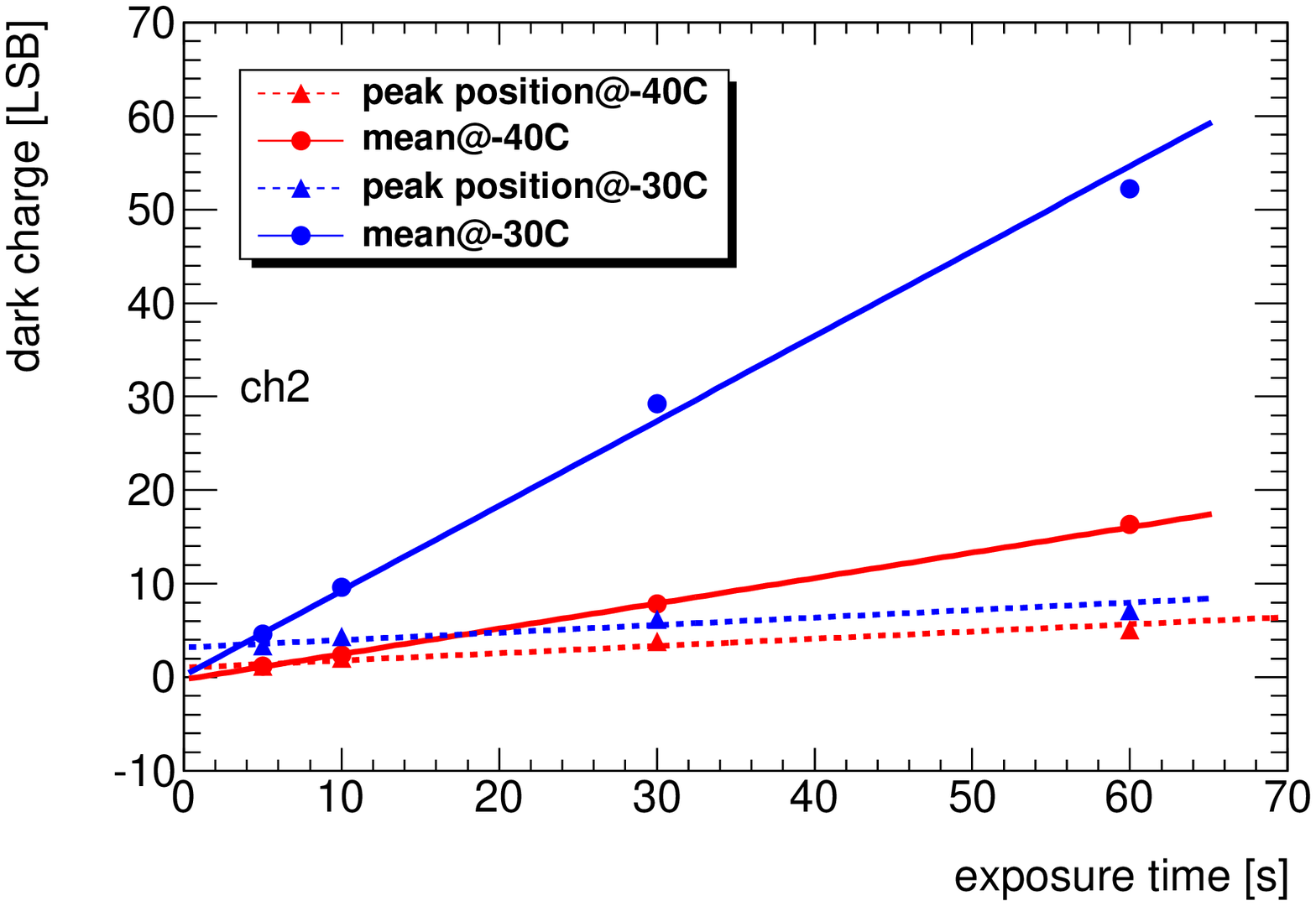}
   \caption{Dark charge as a function of exposure time at  -30 $^\circ$C and -40 $^\circ$C\cite{ref3}}
   \label{figure1}
   \end{minipage}&
	\begin{minipage}[t]{0.45\hsize}
   \centering
   \includegraphics[width=5.5cm]{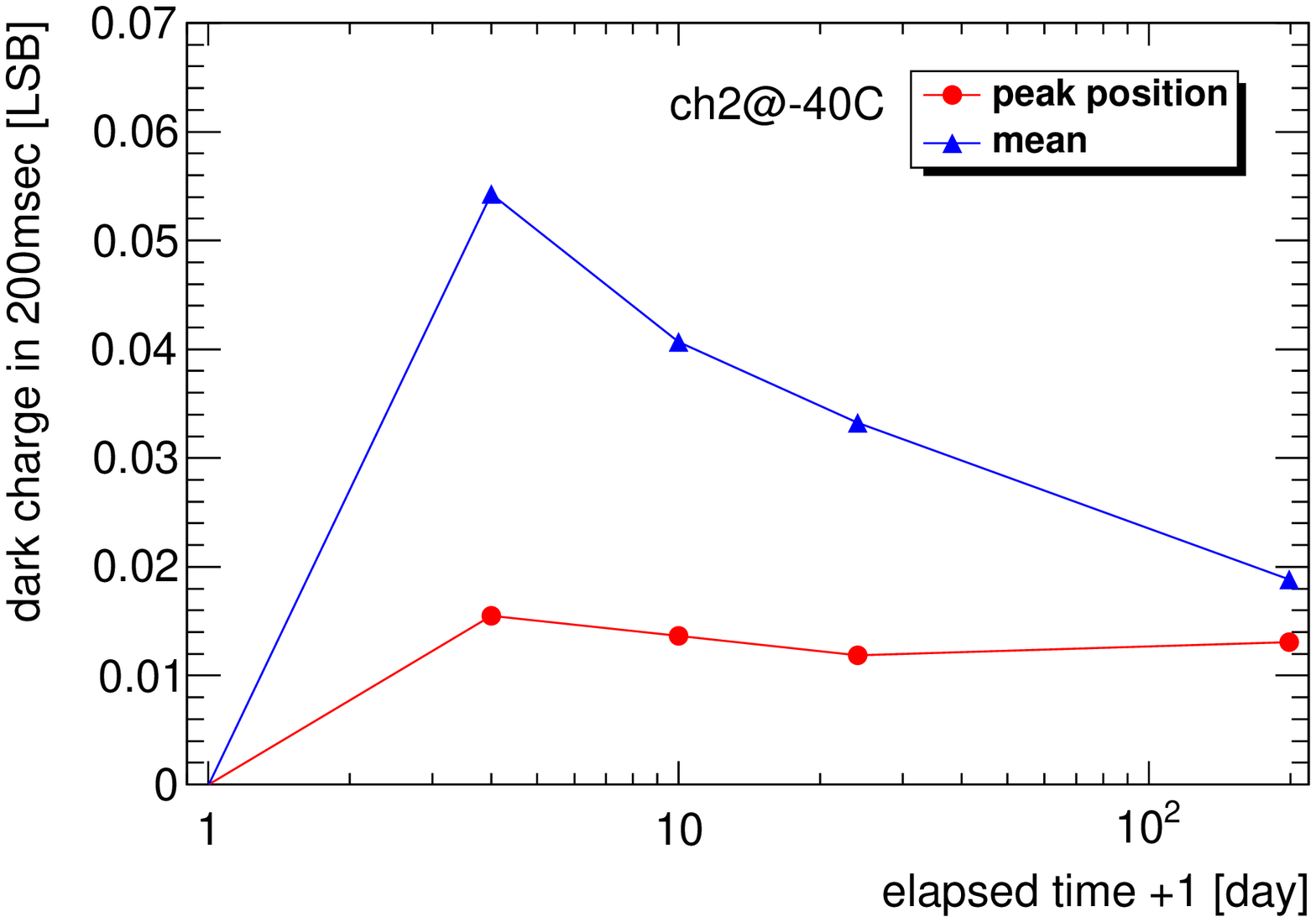}
   \caption{Annealing effect of dark charge\cite{ref3}. The peak position and the mean of dark charge as a function of elapsed day plus 1 day in logalithmic scale are show. Neutron beam was irradiated on 1st day.}
   \label{figure2}
   \end{minipage}
   	\end{tabular}
\end{figure}

The FPCCD detects signal as electrons. However, electorns which come from thermal excitation are detected as dark charge even if there are no signal. Therefore dark charge is accumulated proportional to exposure time of CCD. Dark charge is measured with exposure time 5, 10, 30 and 60 second at -30 $^\circ$C and -40 $^\circ$C. Before the irradiation, dark charge is almost gaussian, but after the irradiation it has a tail. Thus dark charge is measured using two methods. One is the peak position and another is the mean. The peak position is defined by the gaussian component and the mean includes tail effect. Dark charge is proportional to exposure time thus it is fitted with linear function and its slope is dark current(Fig.\ref{figure1}). Dark charge in 200ms, which is time of train gap, is estimated from dark current. Before the irradiation, both indicators, the peak position and the mean, are almost zero. After the irradiation, the dark charge is larger still much smaller than signal charge. Nevertheless, it is enough smaller than signal charge. Annealing effect of the peak position is not seen. For the mean, however, a small annealing effect is seen(Fig.\ref{figure2}). That indicates tail due to the irradiatioin became smaller.

\subsection{Hot pixel fraction}
\begin{figure}
   \centering
   \includegraphics[width=6cm]{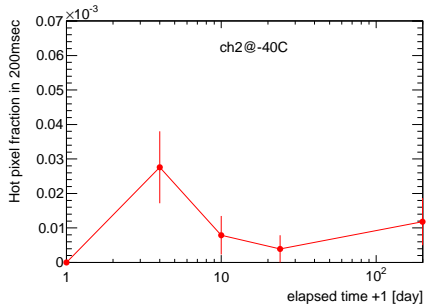}
   \caption{Annealing effect of hot pixel fraction\cite{ref3}. Hot pixel fraction as a function of elapsed day plus 1 day in logalithmic scale are shown. Neutron beam was irradiated on 1st day.}
   \label{figure3}
\end{figure}

There are two kind of hot pixel. One is a hot pixel whose dark charge is always higher than 3 sigma of dark charge. Another is a hot pixel whose dark charge is usually small but sometimes large because of random telegraph noise\cite
{RTS}. Before the irrdiation ADC distribution of dark charge were fitted with gaussian and the mean $(\mathrm{\mu})$ and the width $(\mathrm{\sigma})$ were obtained. We measured dark charge in each pixel many times after the irradiation at -30 $^\circ$C and -40$^\circ$C. When ADC value scaled to 200 msec exposure time is larger than $\mu+3\sigma$ with probability of more than 80\%, we defined it as a hot pixel. Before the irradiation, Hot pixel fraction was 0 but after the irradiation it became $2.76\times10^{-5}$ at -40 $^\circ$C which is much smaller than background occupancy of 2.8\% and requirement of a few percent. Annealing effect is seen and it is consistent with annealing effect of dark charge (Fig.\ref{figure3}). The tail of dark charge comes from hot pixels and hot pixels decreased due to annealing thus tail also decreased.

\subsection{Charge transfer innefficiency}
\begin{figure}
\begin{tabular}{cc}
	\begin{minipage}[t]{0.45\hsize}
   \centering
   \includegraphics[width=6cm]{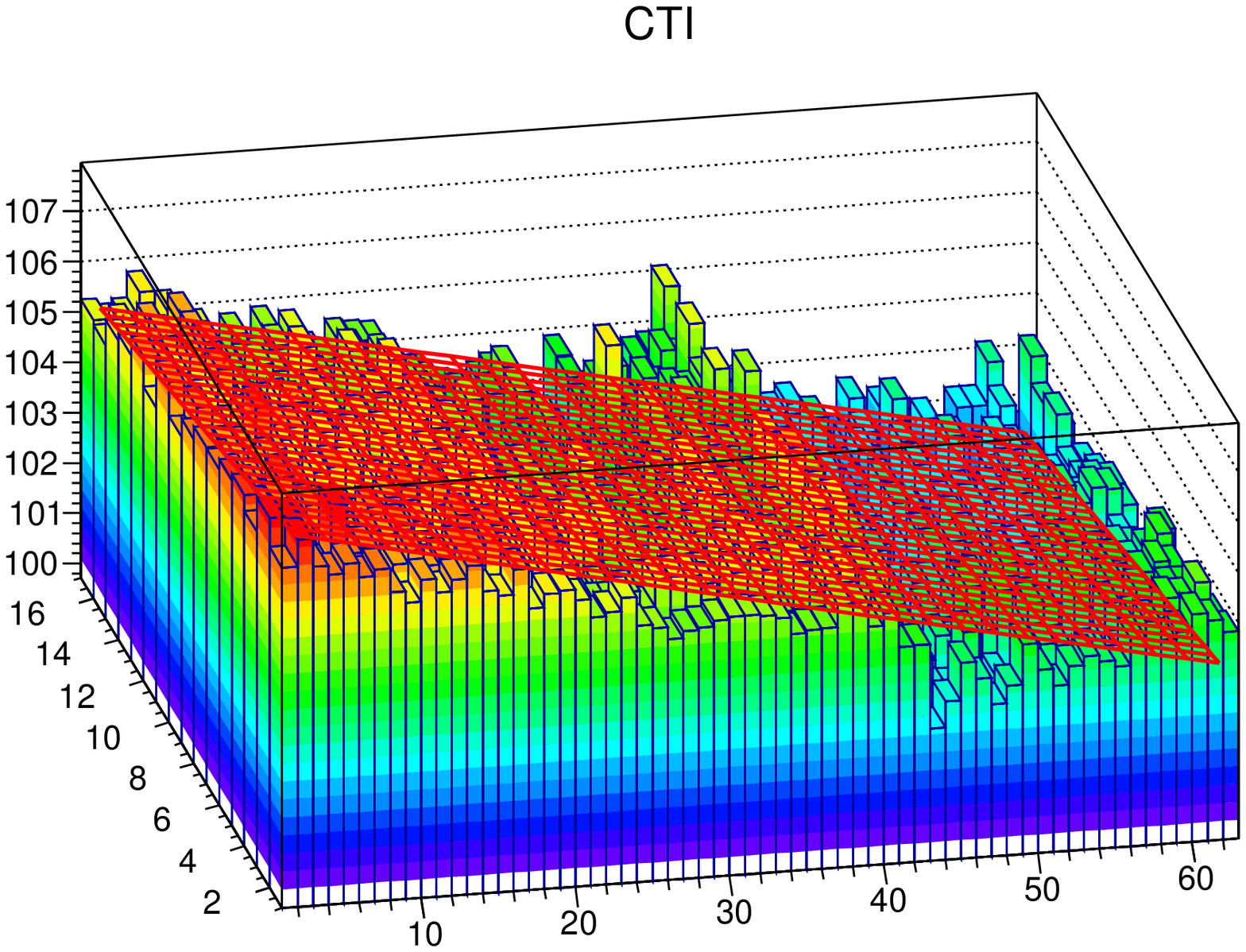}
   \caption{Two dimensional distribution of the peak position (ADC count) of 5.9 keV X-ray from Fe55\cite{ref3}. X and Y ares are horizontal and vertical numbers for super pixel. Readout located at (0,0). The distribution was fitted with the function described in the text.}
   \label{figure4}
   \end{minipage}&
\begin{minipage}[t]{0.45\hsize}
   \centering
   \includegraphics[width=6cm]{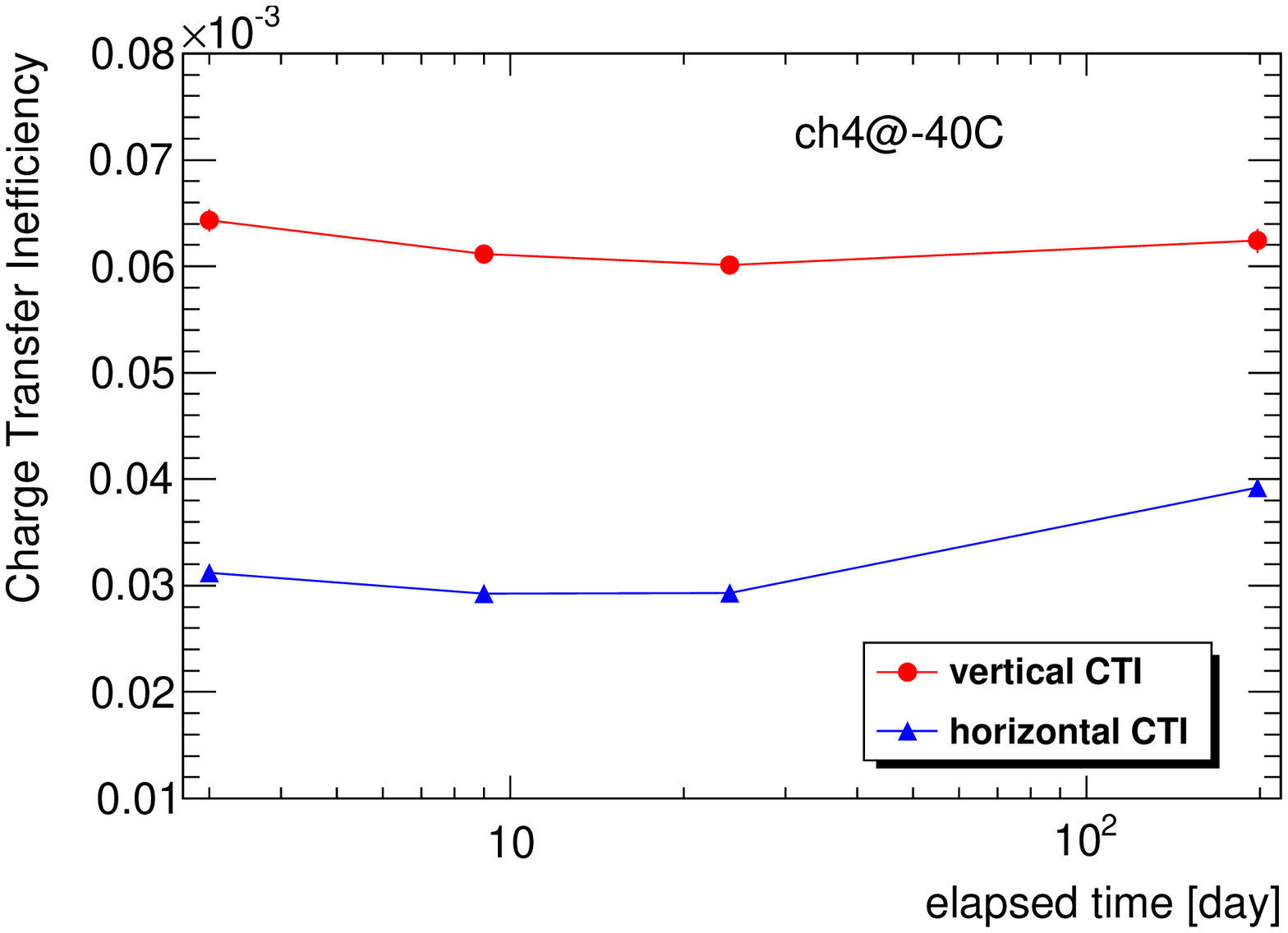}
   \caption{Annealing effect of CTI\cite{ref3}. CTI as a function of elapsed day in logalithmic scale are show. Circle (red) shows vertical CTI and triangle (blue) is horizontal CTI.}
   \label{figure5}
   \end{minipage}
   	\end{tabular}
\end{figure}

CCD transfers signal charge from pixel to pixel to be read out in the end. Ideally charge is transferred completely however charge is actually lost by lattice defect. The main source of lattice defects is radiation damage. Charge transfer inefficiency (CTI) is introduced as an indicator of the charge loss. We difined CTI as inefficiency of one transfer from pixel to pixel. Irradiation by 5.9 keV X-ray from Fe55 is used to measure CTI. Peak position of Fe55 signal in each pixel is fitted with a function of $f(x,y)=S(1-CTI_h)^x(1-CTI_v)^y$ where S is peak position of X-ray from Fe55 before the transfers then CTI is obtained (Fig.\ref{figure4}). Signals are transferred horizontally and vertically, and CTI is defined for each case: $CTI_h$ and $CTI_v$. In this study we used super pixels each of which consists of $16\times16$ pixels instead of pixels because of low statics of X-ray. The CTI's after the irradiation were found as follows: $CTI_h=3.49\times10^{-5}$ and $CTI_v=6.34\times10^{-5}$. A real size of FPCCD will be $13000\times128$ pixels for one readout thus maximum charge loss is $(1-3.49\times10^{-5})^{13000}(1-6.34\times10^{-5})=0.63$. Annealing effect was not seen (Fig.\ref{figure5}). 

\section{Ladder R\&D}
\begin{figure}
\begin{tabular}{cc}
	\begin{minipage}[t]{0.45\hsize}
   \centering
   \includegraphics[width=6cm]{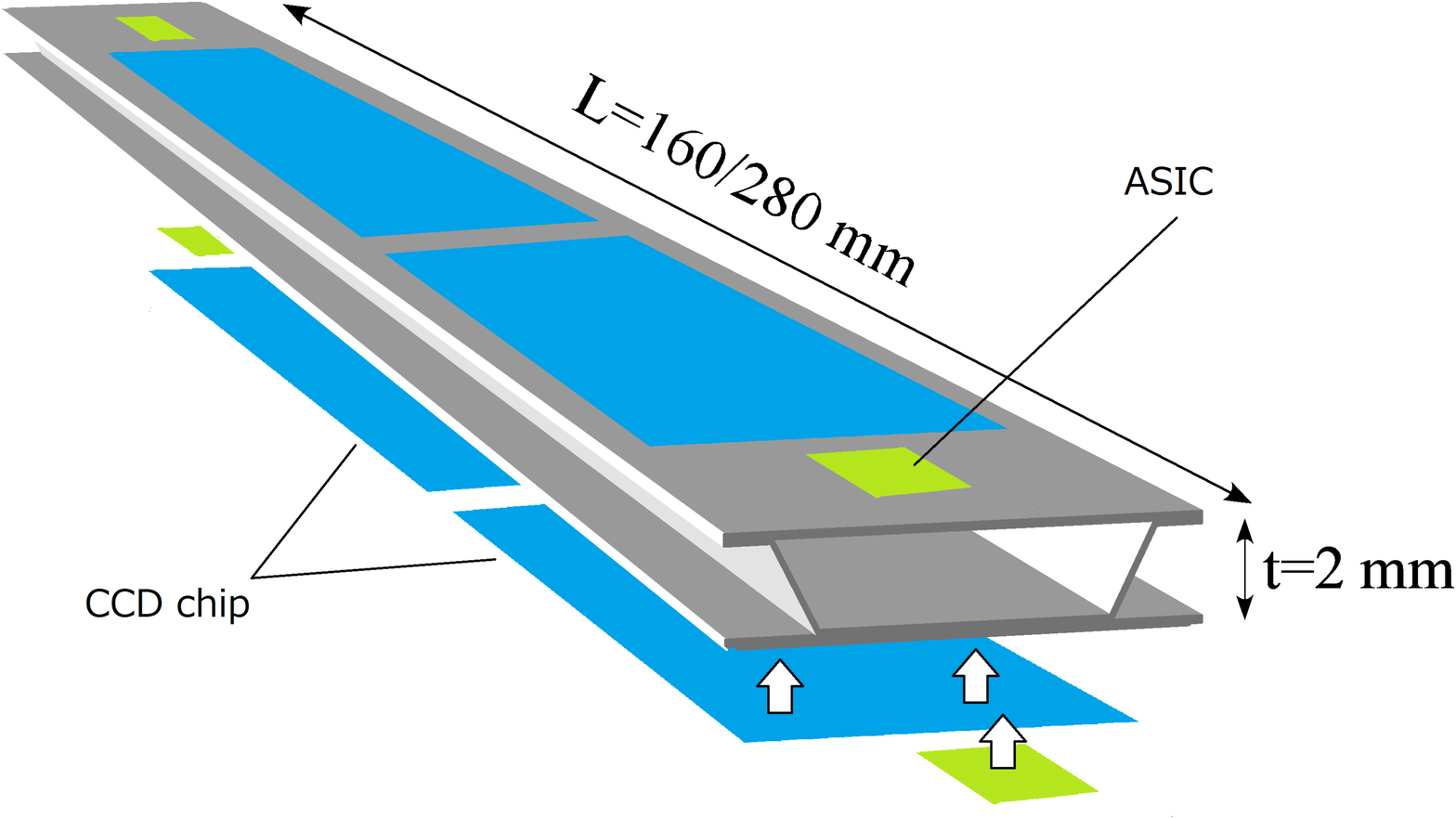}
   \caption{Illustration of ladder design. CCD chips (blue) are attached on both sides and ASICs (green) are attached on both ends.}
   \label{figure6}
   \end{minipage}&
\begin{minipage}[t]{0.45\hsize}
   \centering
   \includegraphics[width=6cm]{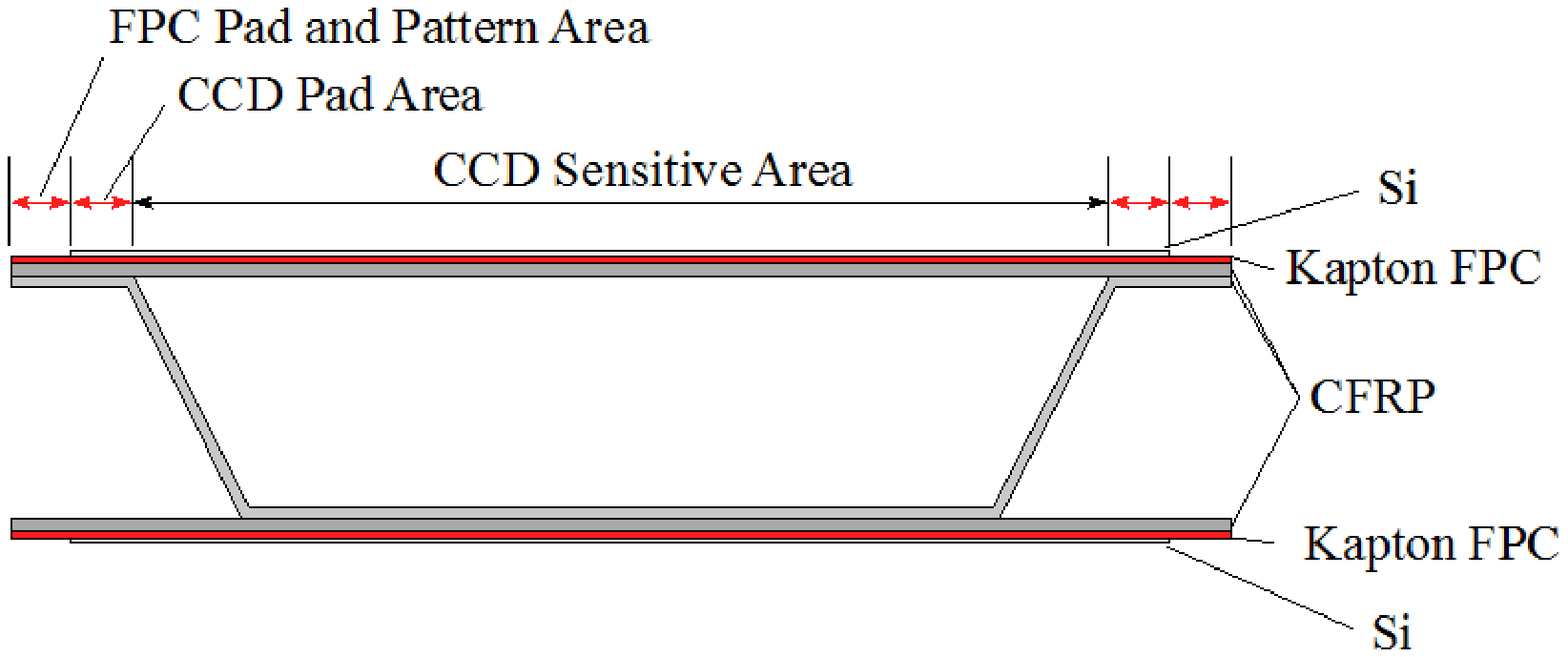}
   \caption{Illustration of ladder structure.}
   \label{figure7}
      \end{minipage}
   	\end{tabular}
\end{figure}

Ladder design idea for the FPCCD is shown in Fig.\ref{figure6}. It has dobble-sided ladder structure and their space is about 2mm. Two FPCCD chips are attached on both sides and two readout ASICs are on both ends. Ladder structure which is made by Si, Kapton FPC and CFRP, is shown in Fig.\ref{figure7}. \par
Silicon wafer is found bending because of thickness which is 50 $\mathrm{\mu}m$ thus gluing by vacuum suction is needed for assembly. Additionally, there are thermal issue. FPCCD will be operated at -40 $^\circ$C. Difference of coefficients of thermal expansion between Si and CFRP is an issue. This stress has to be absorbed by soft glue. R\&D has just begun. 

\section{Two-phase \ce{CO2} cooling system}
FPCCD will be operated at -40 $^\circ$C to improve radiation tolerance. However, space for cooling pipe and thermal insulator inside ILD is very limited. two-phase \ce{CO2} cooling system uses the heat of vaporization to cool detectors which can cool detectors sufficiently thanks to the large latent heat of \ce{CO2} (300 $\mathrm{J/g}$) in spite of the limited space. Therefore two-phase \ce{CO2} is the most suitable choice. There are two options of \ce{CO2} cooling. One is a circulation using liquid pump and another is a circulation using a gas compressor. The latter is our R\&D choice\cite{ref4}. \par
\begin{figure}
   \centering
   \includegraphics[width=9cm]{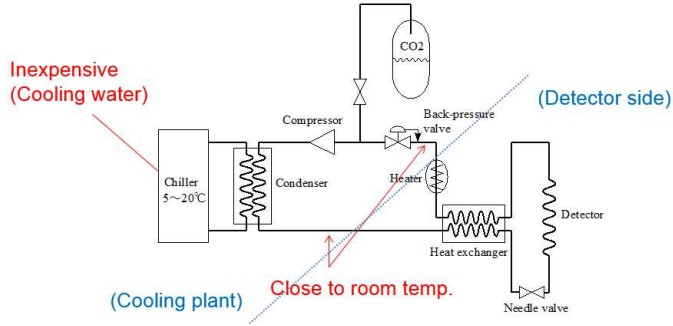}
   \caption{Schematic view of two-phase \ce{CO2} cooling system using a gas compressor.}
   \label{figure8}
\end{figure}

In this cooling plant, \ce{CO2} is circulated near room temperature and cooled by the heat exchanger on the detector side (Fig.\ref{figure8}). Thus a low temperature ( $<-40^\circ$C) chiller and thermal insulator for the long transfer tube between the cooling plant and the detector are not necessary. Cooling water supplied for the ILC experimental hall will be used for \ce{CO2} condensation instead of a chiller. A demerit of this system is need of heater to conpletely vaporize \ce{CO2} returning to gas conpressor.\par
We have demonstrated the cooling between -40 $^\circ$C to +15 $^\circ$C with a prototype cooling system using a gas compressor. 

\section{Summary}
The status of the R\&D studies on the FPCCD vertex detector for the ILC is reported. The neutron irradiation test was performed at CYRIC. An FPCCD sensor was irradiated by neutron of $1.78\times10^{10}\mathrm{(n_{eq}/cm^{2})}$ in two hours, which corresponds to 19 years of ILC running at $\sqrt{s}=500 \mathrm{GeV}$. Dark current and hot pixel fraction get larger after the irradiation; the increase, however, is small enough to be problematic. CTI is also increased but it is acceptable level, as far as neutron background is concerned.\par
Ladder design and assembly ideas are suggested. R\&D has just begun. Two-phase \ce{CO2} cooling system using gas compressor is found to be the most suitable choice for FPCCD. A prototype demonstrated sufficient cooling between -40 $^\circ$C to +15 $^\circ$C.

\end{document}